# Mechanically Tunable Radiative Cooling for Adaptive Thermal Control

*Andrew Butler and Christos Argyropoulos**

Department of Electrical and Computer Engineering, University of Nebraska-Lincoln, Lincoln, Nebraska 68588, USA

**ABSTRACT**

Passive radiative cooling is currently the frontier technology in renewable-energy research. In terms of extraterrestrial applications, radiative cooling is a critical component to the thermal management system of a spacecraft, where the extreme environment of space can cause large temperature variations that can break and damage equipment. For terrestrial applications, nocturnal or daytime radiative cooling is expected to lead to cost-effective passive heat management without the need of inefficient and costly artificial refrigeration technologies. However, most currently available radiative cooling systems cannot be changed dynamically and radiate a constant static amount of thermal power. Dynamically tunable adaptive radiative cooling systems will be a critical development to prolong the lifetime of spacecraft or improve the efficiency of terrestrial cooling systems. Here we propose stretchable radiative cooling designs that can be substantially tuned by using the simple physical mechanism of mechanical strain. When their structure is stretched, the radiated power is significantly reduced. We develop a modeling method that can simulate mechanical stretching combined with electromagnetic response to compute the tunable thermal emission of these new adaptive radiative cooling systems. The presented photonically engineered structures can be used as coatings to achieve efficient adaptive thermal control of various objects in a cost-effective and environmentally friendly way. The proposed designs are much simpler to be realized than others found in the literature and the best design achieves a high thermal emission power with a tunable range on the order of 132 W/m$^2$.

**INTRODUCTION**

Thermal management systems are a critical consideration in the design of various spacecraft components, such as satellites and space shuttles. The harsh environment of space exposes sensitive equipment to an extreme variety of thermal conditions [1]. These harsh conditions can degrade and break electronic equipment and reduce the lifetime of spacecraft. Thermal regulation systems must maintain a stable temperature to avoid damage to electronic equipment [2]. Since transferring heat out of a system through conduction and convection through the air is not possible in space, radiative cooling is the only way to efficiently expel heat from a system. Moreover, nocturnal and daytime passive radiative cooling has been used in a plethora of terrestrial applications leading to cost-effective passive heat management without the need of inefficient and costly artificial refrigeration technologies [3–9]. Interestingly, radiative cooling can be achieved through photonic engineering of the thermal emission of an object. At room temperature (300 K), black body spectral radiance peaks around the wavelength of 10 μm in the mid-infrared (mid-IR) range. Materials and structures with high emissivity in the mid-IR will naturally radiate thermal energy in outer space and passively cool themselves below the room temperature.

The terrestrial or extraterrestrial thermal environment varies greatly and requires thermal management systems to be flexible to respond to a wide range of situations. However, most radiative cooling designs have a static response and are not tunable, i.e., they always radiate the same amount of thermal energy. Typical static radiative cooling designs based on photonically engineered designs and bulk materials include multilayer configurations [4,7,8,10,11], thermally emissive paints and coatings [9,12–14], or more complex photonic structures [15–18]. Once fabricated or applied, these designs are fixed by design and cannot be changed. As an example, thermal management systems in spacecraft need to be able to alternatively heat or cool various equipment. Thus, the constant high radiative cooling will force the system to use more power for heating, which is detrimental to the power-efficiency and lifetime of spacecraft. Dynamically tunable radiative cooling will be an important development for continued smooth operation in space in order to extend the lifetime and improve the durability of equipment in this challenging environment. Similar concerns apply to terrestrial applications, where tunable passive cooling will be beneficial to achieve efficient adaptive thermal control of various objects in a cost-effective and environmentally friendly way.

Currently, the most commonly proposed approach to achieve self-adaptive radiative cooling is to use phase change materials, such as vanadium dioxide ($VO_2$), whose optical properties change at different temperatures [19–21]. At a critical temperature, $VO_2$ changes from a metallic to an insulating state that can be used to change the emissivity from high to low, essentially turning the cooling performance from on to off. While this approach requires no external input to be activated, it operates only as a discrete binary state and the thermal emission can only be at a maximum or a minimum constant value. Moreover, once fabricated, the critical temperature cannot be changed and $VO_2$ radiative cooling designs can only operate near this critical temperature. These factors limit the range of dynamic control that designs based on phase change materials can offer and make them useful only to a few niche applications. Alternative tunable thermal emission devices have been proposed mainly for satellite thermal management, such as electrochromic emissivity modulators that switch emissivity on and off by applying a voltage to the device [22]. However, these designs would be challenging to scale up to larger surface areas and the electrical controls will be difficult to operate in a harsh environment. Finally, another method to achieve emissivity control was recently demonstrated by using porous polymer coatings that change optical properties when wetted by different fluids [23]. While this technique may be viable for tunable radiative cooling on Earth, it is less feasible for extraterrestrial radiative cooling applications where only a limited amount of fluid could be carried on the spacecraft.

One promising alternative way to provide continuous dynamic control of the radiated emission for both terrestrial and extraterrestrial applications is to use stretchable materials that can control the emissivity of a structure through simple mechanical strain. Recently, a stretchable design was theoretically proposed that utilized nanoparticle filled polydimethylsiloxane (PDMS) along with a silver coated PDMS grating [24]. Normally, the embedded nanoparticles and the PDMS give the device high emissivity in the mid-IR range, but when stretched, the emissivity of the device drops because the period of the grating increases. While this design achieved high tunability, the cooling power was relatively low and not appropriate for practical applications. Furthermore, it relies on random nanoparticle volume filling ratios that can be very difficult to be experimentally realized. Another approach was also lately proposed using a multilayered grid structure composed of alternating layers of ethylene propylene diene monomer rubber (EPDM) and air to be used as a stretchable filter for radiative cooling [25]. When stretched, the transmission band of the filter is shifted to the 8-13 μm band in the mid-IR, allowing a thermal emitter in room temperature to radiate through this wavelength range. While this approach of using a stretchable filter is interesting, the required two-dimensional structuring and air gap between the filter and the thermal emitter makes the design difficult to realize in the real world. A similar approach was used for a design consisting of an emissive layer covered by a cracked shield [26]. The cracks in the shield grow wider when stretched and the emissive layer is able to radiate in the surrounding environment. However, this design was not explored for radiative cooling purposes due to its relatively poor emissivity performance in terms of values and tunability. Finally, another relevant design used crumpled graphene on a stretchable substrate to realize tunable mid-IR emissivity [27]. Stretching the structure decreased the pitch of the graphene which reduced the emissivity. Still, the emissivity of this structure in the mid-IR was relatively low and the stretchable graphene requirement increased the design and fabrication complexity.

Given the complexity of other stretchable radiative cooling designs, in this paper we propose much simpler novel stretchable periodic structures to achieve mechanically tunable emissivity and radiative cooling. Our designs consist of periodic mound shaped PDMS embedded with periodic rectangular and cylindrical silicon nitride ($Si_3N_4$) structures. Stretching the PDMS reduces its thickness and increases the periodicity of the PDMS mounds and embedded $Si_3N_4$ structures, which leads to reduced absorption and emissivity of the entire structure in the mid-IR range. Additionally, we show how a previously demonstrated crumpled metallic layer can be used as a substrate to further improve the tunability of the proposed radiative cooling designs [28]. By controlling the strain of the structure, the emissivity can be dynamically tuned over a wide range of values. We develop a modeling method that can simulate mechanical stretching combined with electromagnetic response to compute the tunable thermal emission of these new radiative cooling systems. The presented photonically engineered adaptive structures will achieve tunable passive radiative cooling for terrestrial and extraterrestrial applications.

**THEORETICAL DESIGN**

PDMS is a unique and widely used material with high elasticity and low Young's modulus, meaning it can be extensively stretched and deformed relative to its size [29]. Our designs consist of periodic mounds of PDMS with periodicity $p_{PDMS_0} = 45\mu m$ and a maximum thickness $t_{curve_0} = 2.5\mu m$ with unit cells shown in Figure 1. The mounds rest on a continuous flat rectangular-shaped PDMS piece with a thickness $t_{flat_0} = 1.5\mu m$. Embedded within the PDMS are simple periodic $Si_3N_4$ geometries. The first design uses rectangular strips of $Si_3N_4$ with equal width $w = 6.5\mu m$, thickness $t_{Si3N4} = 0.9\mu m$, and period $p_{rect_0} = 9\mu m$. The top edges of the rectangular strips are offset above the top edge of the flat PDMS by $f_{rect} = 300nm$ (Figure 1a). Our second design uses cylindrical $Si_3N_4$ rods with radius $r = 350nm$ and a period $p_{cyl_0} = 0.75\mu m$. The centers of the cylinders are offset below the top edge of the flat PDMS by $f_{cyl} = 100nm$ (Figure 1b). The purpose of these offsets is to prevent the $Si_3N_4$ structures from breaking the PDMS during the stretching process. They are defined as the height of the $Si_3N_4$ structures protruding above the top edge of the flat PDMS.

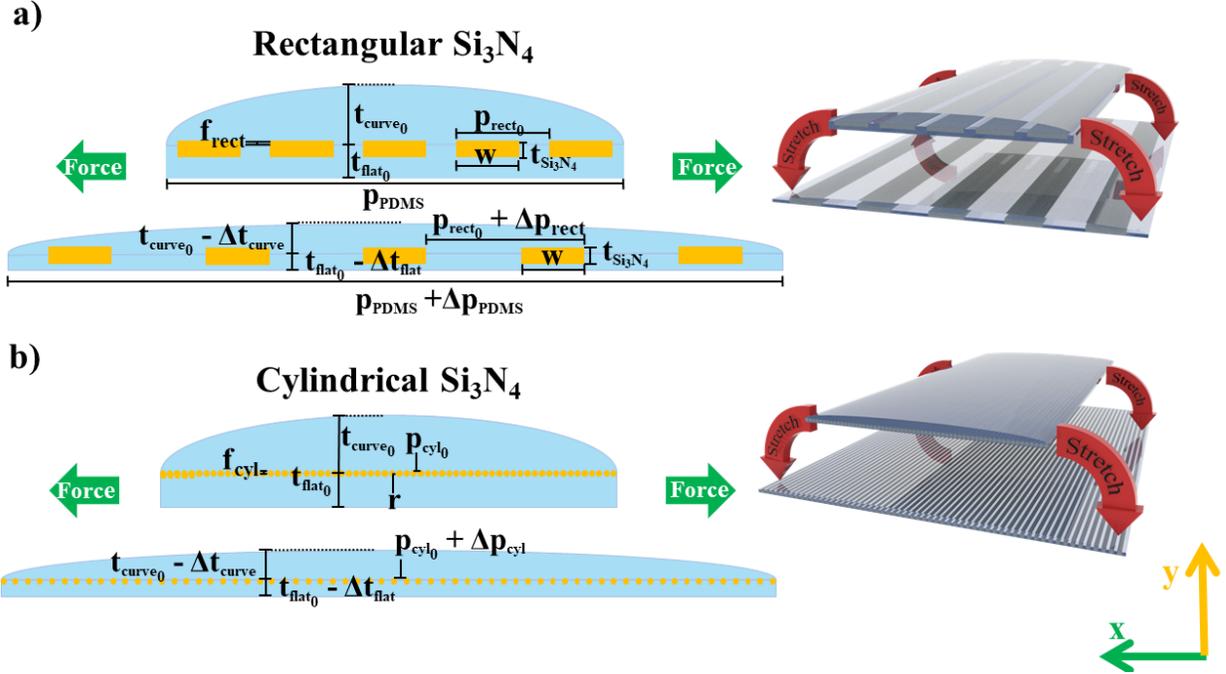

**Figure 1.** Schematic of the unit cell of two stretchable radiative cooling designs based on PDMS embedded with a) rectangular and b) cylindrical $Si_3N_4$ structures.

As shown in Figure 1, when a lateral force along the x-direction is applied to the PDMS, the periodic structure is stretched and undergoes strain and deformation. Strain is defined as the ratio between the deformation $\Delta x$ and the original length $x_0$ of the structure: $\varepsilon = \frac{\Delta x}{x_0}$ [30]. Under strain, the periodicities of both the PDMS and $Si_3N_4$ structures are increased according to $p = p_0 + \Delta p$, where $p_0$ is the initial unstretched period and $\Delta p = \varepsilon * p_0$ is the strain-induced increase in the periodicity. However, the size of the $Si_3N_4$ structures ($w$, $t_{Si3N4}$, and $r$) do not change during the stretch process because $Si_3N_4$ has a much higher Young's Modulus of 280 GPa [31]. The $Si_3N_4$ structures remain embedded in the PDMS but move farther apart as the structure is stretched. The x-direction stretching also causes deformation and strain along the vertical y-direction. The ratio of the strain in the y- to the x-direction is defined as Poisson's Ratio: $\nu = -\frac{\varepsilon y}{\varepsilon x}$ [30]. Here, the minus sign indicates that the structure is shrinking in the vertical y-direction. Values for the Poisson's Ratio of PDMS range in the literature from 0.45 to 0.5 [32]. For our purposes, we choose an intermediate value of: $\nu = 0.48$ [33]. Thus, the thickness of the PDMS shrinks according to $t = t_0 + \Delta t$, where $t_0$ is the initial unstretched thickness and $\Delta t = \varepsilon * \nu * t_0$ is the decrease in thickness due to the lateral strain. This process is similar to mechanically tunable photonic crystals, where a periodic array of silicon (Si) pillars embedded within PDMS was used [34]. When stretched, the distance between the pillars increased laterally (same direction as strain) and decreased axially (perpendicular to strain). A similar design was experimentally demonstrated using vertically aligned carbon nanotubes embedded in PDMS that was used as a strain sensor [35]. Since the dimensions of the $Si_3N_4$ structures are not changed by the stretching, the Poisson's Ratio of $Si_3N_4$ is not considered in our calculations.

The currently proposed designs are practical, since they can be fabricated by using a molding process similar to a previously demonstrated silver nanowire-embedded PDMS electrode design or a wearable strain sensor based on the PDMS-embedded conductive fabric created in [36,37]. The rectangular $Si_3N_4$ strips can be produced using femtosecond laser micromachining or lithography [38]. The cylindrical structures can be grown in a similar manner to previously reported silicon nitride nanofibers with rectangular cross section dimensions exceeding 1μm, grown using a carbothermal reduction process [39]. Note that the growth of ultralong $Si_3N_4$ nanowires with maximum diameters of 960nm has also been reported as well as the production of aligned ultralong $Si_3N_4$ nanowires with 200nm diameters [40,41]. Once the $Si_3N_4$ structures are fabricated, they could be combined with liquid PDMS in a mold to form the currently presented tunable radiative cooling designs.

In the designs presented thus far, the emissivity reduction is mainly achieved by increasing the transmittance. Recently, a triple mode device that changes between an emissive, reflective, and transparent state in the mid-IR region through stretching was demonstrated [28]. The design consisted of a crumpled metallic substrate and a polymer superstrate composed of a styrene-ethylene-butylene-styrene (SEBS) material. When stretched, the metallic layer flattens, and the device becomes predominantly reflective in the mid-IR range. Further stretching of the device causes the metal to break apart and the device becomes primarily transparent. The process is reversible but has only been explored in the general applications of radiative heat management and thermal camouflage, i.e., not on the emerging technology of passive radiative cooling.

Here, we also explore an approach where the transmission becomes zero for both stretched and unstretched cases when a crumpled metallic substrate is introduced to the designs of Figure 1 [28]. In this scenario, the emissivity of the stretched structures can be further reduced by increasing their reflectance. The metallic designs are shown in Figure 2 for both cylindrical and rectangular $Si_3N_4$ inclusions. We use a 300 nm thick crumpled gold layer with a periodicity $p_{Au} = 1.25$ μm and an amplitude $a_{Au} = 250\ nm$. However, even thinner or thicker crumpled gold layer dimensions with a range of $\pm 5 \mu m$ or even more will also lead to similar response. The crumpled metal is modelled as a sinusoidal geometry with the center of the shape positioned 1μm below the edge of the PDMS flat part. This geometry could realistically be fabricated using a mold created from Laser Induced Periodic Surface Structures (LIPSS). LIPSS structures are micro/nanoscale ripples that form on the surface of materials (usually metals) processed by linearly polarized laser irradiation [42]. The dimensions of the surface ripples can be controlled by changing the laser parameters such as wavelength, pulse count, pulse duration, and fluence. Using LIPSS ripples as a mold, the bottom surface of the PDMS will be textured with the sinusoidal pattern. The PDMS will then be stretched until the bottom surface is nearly flat, and the Au layer will be deposited onto this surface. When the PDMS is unstretched, the Au layer will crumple into the required sinusoidal shape. Metal films have been reportedly grown on PDMS substrates using DC magnetron sputtering [43], thus the proposed designs are feasible to fabricate.

The crumpled metal prevents transmission through the structure independent of the stretching condition. Since the metallic layer is attached to the PDMS, the change in dimensions caused by the stretching is modeled the same way as before. More specifically, the strain is used to determine the change in the periodicity and the Poisson's Ratio of PDMS to determine the reduction in the amplitude of the metallic crumples. Similar to $Si_3N_4$, the thickness of the gold is unchanged by the stretching and only its lateral geometry varies due to its attachment with the elastic PDMS.

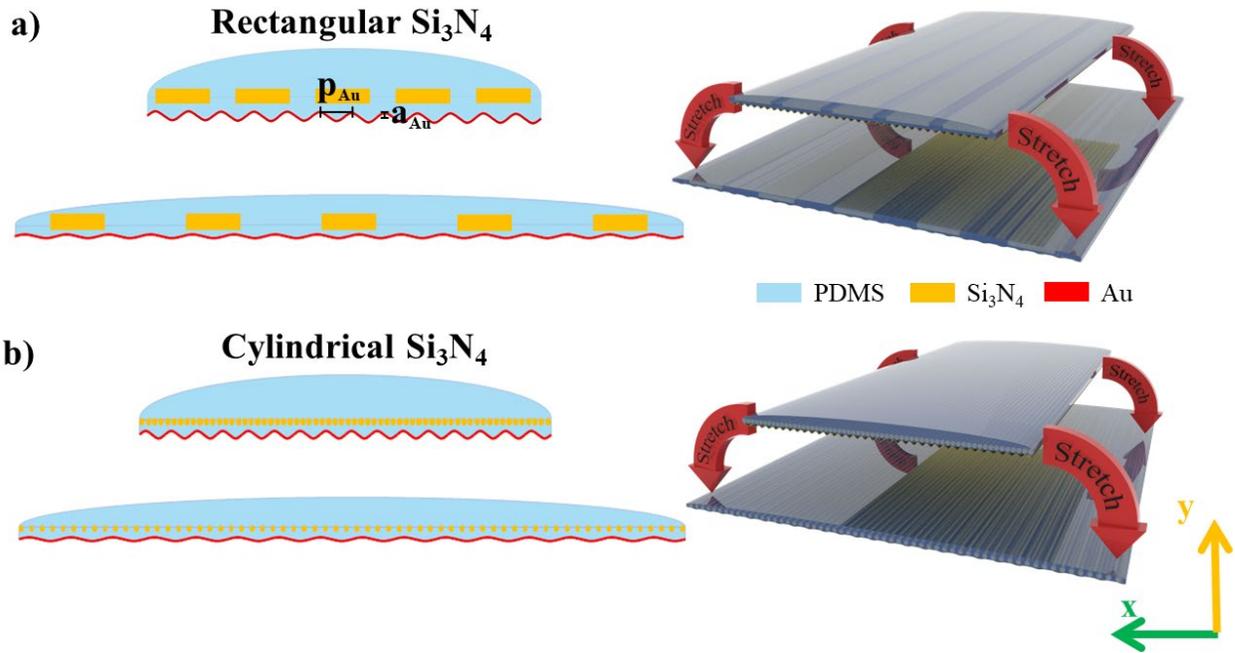

**Figure 2.** Schematic of the unit cell of two stretchable radiative cooling designs based on PDMS over crumpled metal embedded with a) rectangular and b) cylindrical $Si_3N_4$ structures.

The reflection of the metal increases the absorption of the PDMS by effectively doubling the thickness seen by the reflected light. To counteract this detrimental effect, the thicknesses of the PDMS layers and the $Si_3N_4$ structures are reduced compared to the previous designs. The new dimensions are $t_{curve_0} = 1.5 \mu m$, $t_{Si_3N_4} = 600nm$, and $r = 180nm$ for rectangular and cylindrical Si3N4 structures, respectively. The decreased thickness leads to an even more compact tunable radiative cooling design. The positions of the Si3N4 structures are also shifted so that the rectangular structures are centered on the top of the flat section of PDMS, and the center of the cylindrical structures are positioned 120nm below the top of the flat PDMS. The new designs are simulated using the same method as before and the gold material is modeled employing its frequency dependent refractive index [44].

**RESULTS AND DISCUSSION**

We assume that the structures shown in Fig. 1 are much larger in the z-direction compared to the emitted wavelength. As a result, two-dimensional simulations are required to compute their emissivity spectra by calculating their reflectance and transmittance parameters. The absorptivity $A$ of the structures can be computed by using the formula: $A = 1 - |S_{11}|^2 - |S_{21}|^2$, where $S_{11}$ is the reflection coefficient and $S_{21}$ is the transmission coefficient. The absorptivity is equal to the emissivity $E = A$ in the far field due to the Kirchhoff's law of thermal equilibrium. The produced thermal radiation is an incoherent source of infrared light; thus its polarization is random. To account for this important issue in all our calculations, the spectral and angular emissivity is computed as the average of the emissivity for both transverse electric (TE) and transverse magnetic (TM) polarizations. In addition, the structures were simulated by using periodic boundary conditions on the curved PDMS sides, where each unit cell is schematically depicted in Figure 1. The experimentally obtained realistic data of the frequency dependent permittivity of PDMS and Si3N4 were used in all simulations [45,46]. More details on the simulations are included in the Supplementary Information [47]. The developed theoretical model can be applied to any tunable thermal emitter based on mechanically stretchable materials, since it can accurately calculate mechanical stretching and its effect in the electromagnetic and thermal response. Table 1 summarizes the mechanical and electromagnetic parameters of the materials used in our modeling.

**Table 1.** Mechanical and electromagnetic material parameters used in our modeling.

|  | **PDMS** | **$Si_3N_4$** |
|---|---|---|
| Young's Modulus | 0.57MPa – 3.7MPa [29] | 280GPa [31] |
| Poisson's Ratio | 0.45 – 0.5 [32,33] | 0.23 [48] |
| Peak Extinction Coefficient | ~0.72 @ 12.5um [46] | ~2.3 @ 11um [45] |

PDMS is mostly transparent to light in the visible and near infrared spectrum, but the imaginary part of its refractive index becomes significant for wavelengths above 6 μm due to phonon resonances [46]. This coincides with the peak of room temperature (300K) blackbody radiation, and PDMS can radiate thermal energy in the mid-IR atmospheric transparency window. The computed emissivity of our PDMS periodic structure without any $Si_3N_4$ inclusions, i.e., simplest possible configuration, is shown in Fig. 3a for both the unstretched and stretched cases. While the emissivity is high in some parts of the mid-IR spectrum, there are substantial dips especially around 11 μm which turn the average emission of this structure rather low but moderately tunable, as can be seen in the angular emissivity plots also shown in Figure 3a (lower caption) where the average value of emissivity was computed along the atmospheric window (9-13 μm) for different emission angles. In the stretched case, a strain of 120% was used, which is the maximum possible value that can be achieved by PDMS without the occurrence of cracks [24]. Thus, the demonstrated results are the maximum and minimum possible emissivity spectra for our structure under stretching and define the maximum possible tunable thermal radiation emission range. The main mechanism causing the decrease in emissivity between the unstretched and stretched cases is the deformation of the structure geometry, especially in the y-direction (height). The structure becomes wider and thinner, decreasing its absorption and interaction with light.

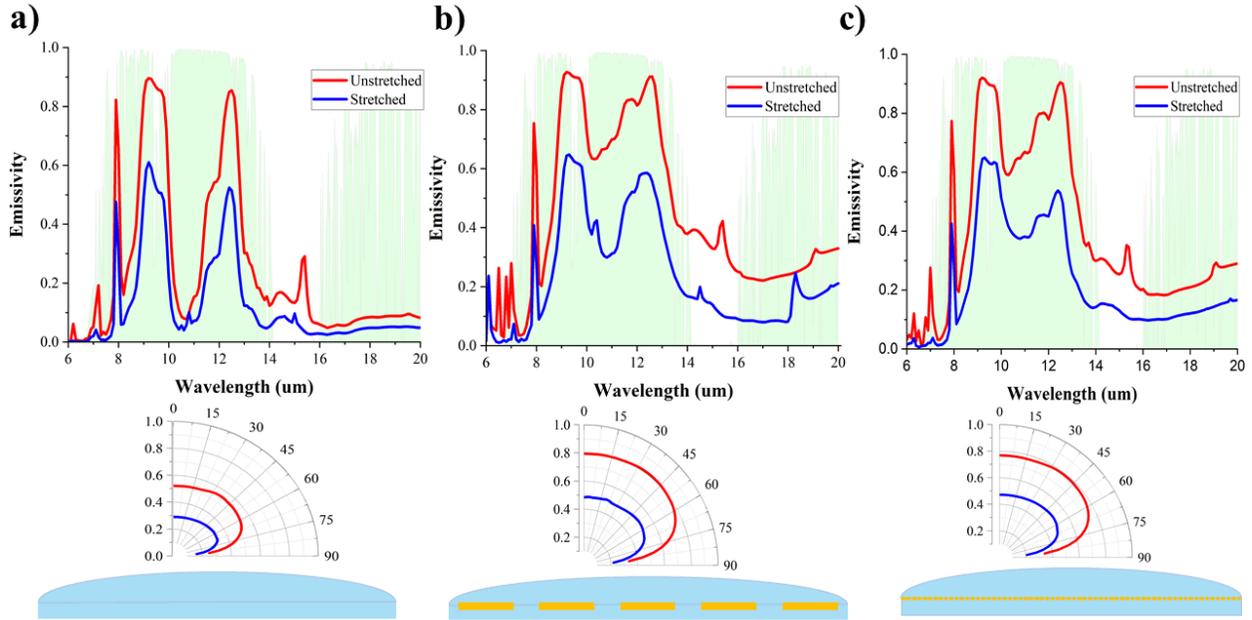

**Figure 3.** Emissivity spectra and angular emissivity of the PDMS structure a) without $Si_3N_4$ inclusions, b) with rectangular $Si_3N_4$, and c) with cylindrical $Si_3N_4$. The green shaded area represents the atmospheric transmission window. Lower captions: angular emissivity is the average emissivity value computed over 9-13 μm for different emission angles.

The tunability can be improved by the inclusion of $Si_3N_4$ structures inside the PDMS design. The imaginary part of the refractive index of $Si_3N_4$ peaks near 11.5 μm, the same wavelength where the absorption of PDMS significantly decreases [45]. Note that simply adding a layer of $Si_3N_4$ to the bottom of the PDMS is not possible because of $Si_3N_4$ high Young's Modulus. Hence, $Si_3N_4$ cannot stretch as far as PDMS without cracking and fracturing. Therefore, our design uses non-continuous periodic $Si_3N_4$ structures that can move within the PDMS without stretching. The computed emissivities of the proposed designs with $Si_3N_4$ rectangular and cylindrical structures embedded in PDMS are shown in Figures 3b and 3c, respectively, and their average angular emissivity in the range λ = 9-13 μm is shown in the lower insets.

Interestingly, the shape of the $Si_3N_4$ structures has a pronounced effect on the total emissivity values and their tunable range. The rectangular design has slightly higher emissivity than the cylindrical design in both the stretched and unstretched cases, but the tunability in the cylindrical design is improved since the difference between the stretched and unstretched cases is greater. The thickness, width, and spacing of the structures affect the tunability and the final designs are optimized given that they provided the best results out of a wide variety of sizes and shapes. The main mechanism that causes the increase in tunability near λ = 11 μm is the change in periodicity between the stretched and unstretched configurations. When the gaps between the $Si_3N_4$ structures are increased, more light is able to pass unaffected through the structure and the absorption/emissivity is reduced. The physical mechanisms that enable the large differences in mid-IR emissivity between the stretched and unstretched states are further elucidated by the electric field enhancement distributions plotted in the Supplementary Information [47].

To determine the cooling performance for applications on satellites or other space applications, only the thermal power radiated by the structure $P_{rad}$ needs to be considered. Hence, $P_{rad}$ can be calculated by using the formula:

$$P_{rad}(T) = 2\pi A \int_0^{\pi/2} \sin(\theta)\cos(\theta)d\theta \int_0^{\infty} I_{BB}(T,\lambda)\mathrm{E}(\lambda,\theta)d\lambda, \qquad (1)$$

where A is the surface area of the structure, θ is the direction of emission, T is the temperature, and E(λ,θ) is the spectral and angular emissivity computed in the previous section. The blackbody spectral radiance $I_{BB}$ is calculated through Planck's Law:

$$I_{BB}(T, \lambda) = (2hc^2/\lambda^5)(exp[hc/\lambda k_B T] - 1)^{-1}, \qquad (2)$$

where *h* is Planck's constant, $k_B$ is Boltzmann's constant, *c* is the speed of light, and *T* is the temperature. Figure 4 shows the radiated cooling power versus temperature for our three designs when the goal is to keep the electronics and other satellite equipment at moderate temperatures close to room temperature (300 K) so they will not be damaged but at the same time be able to demonstrate the tuning in the temperature range. Note that larger tuning and higher radiative cooling power values can be achieved in higher temperatures.

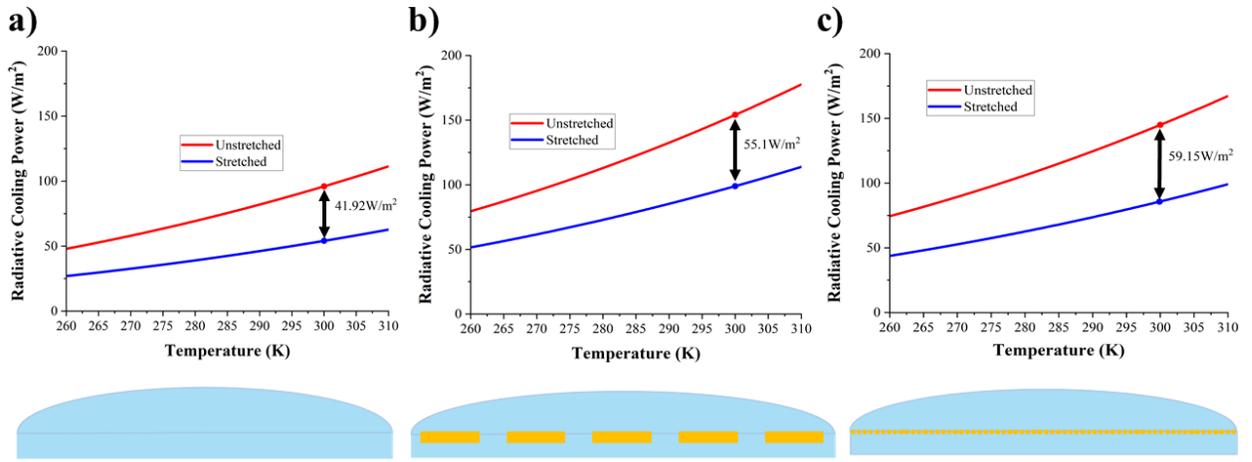

**Figure 4.** Radiated cooling power versus temperature for the PDMS structure a) without $Si_3N_4$ inclusions, b) with rectangular $Si_3N_4$, and c) with cylindrical $Si_3N_4$.

The cooling power at room temperature (300 K) achieved by the rectangular $Si_3N_4$ design ranges from ~98.95W/m² to ~154.1W/m². For the cylindrical design, the cooling power is slightly lower and ranges from ~85.8W/m² to ~144.95W/m². However, the cylindrical design has better tunability with a range of 59.15W/m² which is larger than the rectangular design with a tunable range of 55.1W/m². However, the rectangular design is able to cool to lower temperatures than the cylindrical design in both the stretched and unstretched cases. Without the $Si_3N_4$ structures, the tunability and especially the radiated power is substantially decreased.

Our design can also be used for terrestrial nocturnal tunable radiative cooling applications. To calculate the cooling power on Earth during the night, thermal emission from the atmosphere $P_{atm}$ and conductive and convective heat sources $P_{con}$ should be considered in the power calculations. Taking into account these additional heat sources, the cooling power $P_{cool}$ can be calculated by using the relation:

$$P_{cool}(T) = P_{rad}(T) - P_{atm}(T_{amb}) - P_{con}(T). \tag{4}$$

The absorbed power from the atmospheric thermal emission can be calculated as:

$$P_{atm}(T_{amb}) = 2\pi A \int_0^{\pi/2} \sin(\theta)\cos(\theta)d\theta \int_0^\infty I_{BB}(T_{amb}, \lambda) E_{atm}(\lambda, \theta) E(\lambda, \theta) d\lambda, \tag{5}$$

where $T_{amb}$ is the ambient room temperature (300 K) and $E_{atm}(\lambda, \theta)$ is the atmospheric spectral and angular emissivity that can be calculated from the atmospheric transmittance $t(\lambda)$ by using the relationship: $E_{atm}(\lambda, \theta) = 1 - t(\lambda)^{1/\cos(\theta)}$ [49–52]. Finally, the heating from conductive and convective sources in the environment can be computed by:

$$P_{con}(T, T_{amb}) = Ah_{con}(T_{amb} - T), \tag{6}$$

where $h_{con}$ is the heat transfer coefficient. For the current calculations, we use a value of $10W/m^2/K$ which is typical for ambient air convection [53]. Figure 5 demonstrates the induced cooling temperatures of the currently proposed structures over the course of nighttime. These plots were generated by calculating the equilibrium temperature of our structure given the ambient air temperature during the same time. The ambient temperature data was taken as the recorded temperatures for Omaha Nebraska on June 27-28, 2021 [54].

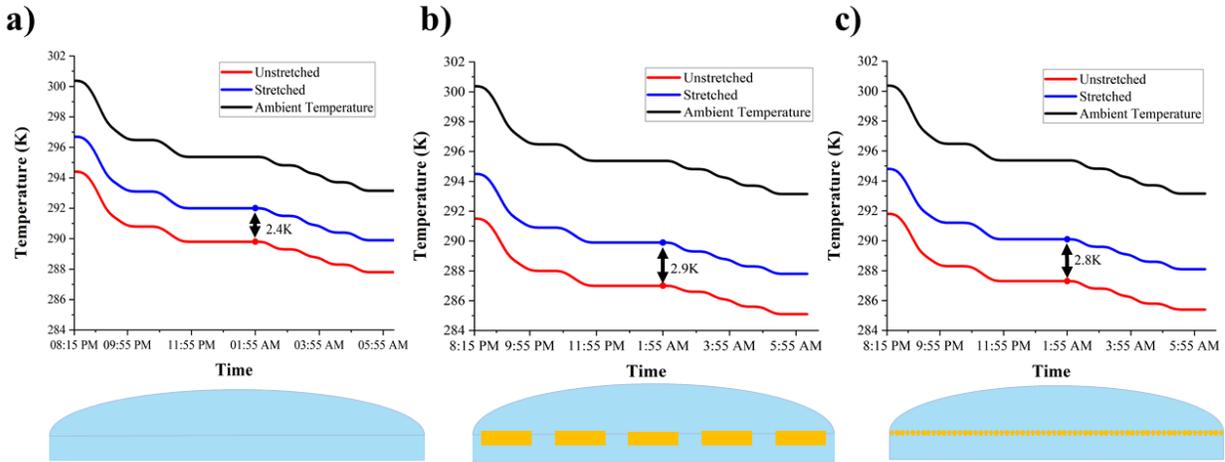

**Figure 5.** Computed hourly temperature plots for the PDMS structure a) without $Si_3N_4$ inclusions, b) with rectangular $Si_3N_4$, and c) with cylindrical $Si_3N_4$ in a terrestrial nocturnal radiative cooling scenario.

The inclusion of the atmospheric emission and conductive and convective sources reduces the cooling performance of our structures compared to the satellite application. However, our tunable structures are still able to cool several Kelvin below ambient temperatures and the difference between the stretched and unstretched cases is clearly illustrated in Figure 5. The currently presented adaptive radiative cooling designs have high and tunable emissivity values and are able to substantially cool below ambient, especially in the case of $Si_3N_4$ inclusions.

The tunability and cooling performance can be further improved by the inclusion of the metallic substrates (shown in Figure 2). The emissivity spectra of the stretched and unstretched metallic designs for normal incidence are shown in the upper captions of Figure 6. The addition of the crumpled metal substantially increases the difference in emissivity between the stretched and unstretched states with and without the $Si_3N_4$ inclusions. As the metal layer is stretched, the height of the crumples is reduced and the layer becomes flatter, leading to a substantial increase in the reflectance. At the same time, the reduction in thickness of the PDMS and increase in periodicity of the $Si_3N_4$ strips further reduce the total absorption of the structure. Interestingly, the average angular emissivity for the atmospheric window range $\lambda$ = 9-13 μm (shown in the lower insets of Figure 6) is significantly improved compared to the designs presented in the previous section. In particular, the PDMS over crumpled metal design with rectangular $Si_3N_4$ inclusions (middle lower caption of Fig. 6) achieves an almost perfect broadband omnidirectional emission in the unstretched case which is considerably reduced in the stretched design leading to an unprecedented mechanically tunable performance compared to relevant designs [55]. Interestingly, the differences in the results of the rectangular versus the cylindrical structures in Figures 6(b) and 6(c), respectively, are mainly due to the large difference in size (area) of the $Si_3N_4$ structures and not due to their shape, as it is demonstrated in Figure S6 in the Supplementary Information [47]. We also checked the effect of the size and fill factors of the $Si_3N_4$ structures, and the results are shown in Figures S7-S9 in the Supplementary Information [47]. Overall, the variation in the emissivity with the $Si_3N_4$ dimensions is small and it can be concluded that the designs are generally robust against fabrication imperfections. In addition, it can be derived that the average emissivity will slightly change with the fill factor once the fill factor increases beyond a certain point. This can correspond to potential nonuniform strain along the x-direction in PDMS, where the main effect will be that the fill factor would change from unit cell to unit cell. Since the effect of the fill factor is small, we can surmise that the potential effect of non-uniform strain on the device performance is also expected to be insignificant.

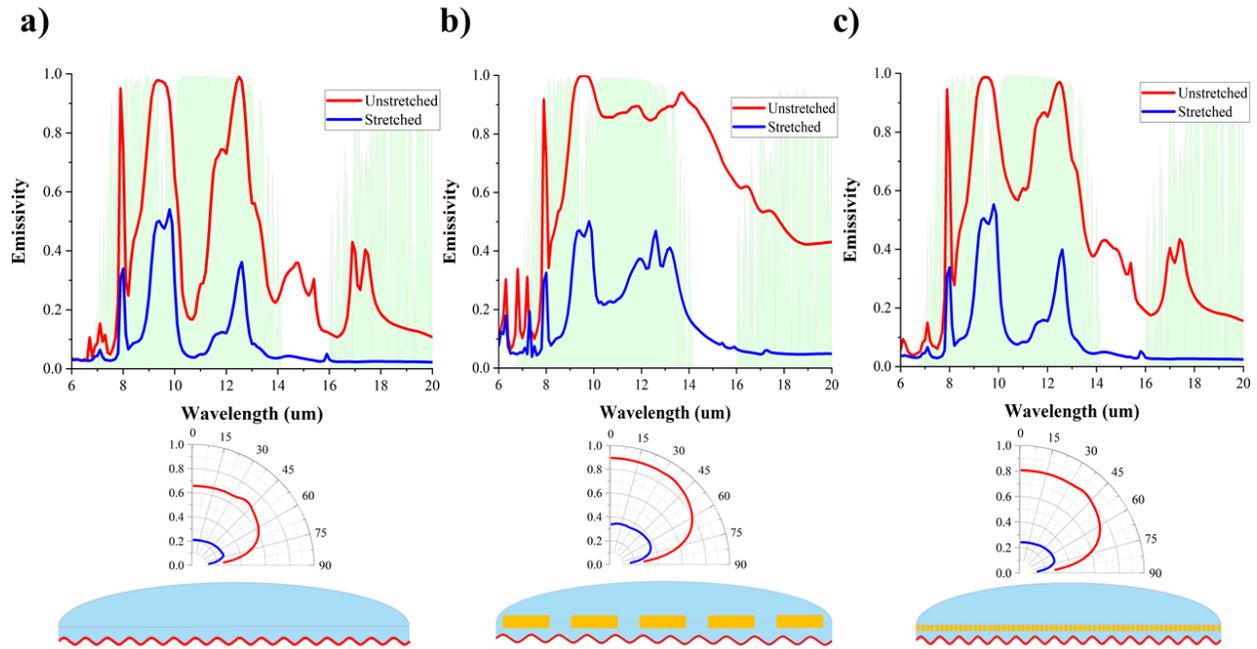

**Figure 6:** Emissivity spectra and angular emissivity of the PDMS over crumpled metal designs a) without Si$_3$N$_4$ inclusions, b) with rectangular Si$_3$N$_4$, and c) with cylindrical Si$_3$N$_4$. The green shaded area represents the atmospheric transmission window. Lower captions: angular emissivity is the average value computed over 9-13 μm for different emission angles.

Figure 7 shows the radiated power of these new designs versus temperature for applications on satellites or other space applications. Moreover, Figure 8 demonstrates the temperature versus time for these devices in a nocturnal radiative cooling scenario on Earth. The findings in Figures 7-8 are computed in a similar way to the results in the previous section. The unstretched structure with rectangular Si$_3$N$_4$ inclusions radiates ~199.5 W/m$^2$ at 300K, as shown in Figure 7, while the stretched structure radiates ~67.56 W/m$^2$, providing an extremely broad tunable thermal emission range of ~131.94W/m$^2$. This large tunability is further illustrated by the nocturnal radiative cooling application on Earth where there is a temperature difference of ~5.1K between the stretched and unstretched states for the same design (middle caption in Figure 8). As before, the tunability and radiated power are significantly reduced without the embedded $Si_3N_4$ structures. Moreover, the addition of the crumpled metal substrate further improves the tunability compared to the previous section designs and prevents transmission of IR light through the structure. This substantial improvement in tunability is due to the increase in reflectance caused by stretching the metal that eventually becomes flatter. Although the tunable temperature range of the rectangular $Si_3N_4$ design in Figure 8 is only 0.8K larger than the case without the $Si_3N_4$ inclusions, the tunable range of the radiated power in Figure 7 is much larger, hinting that the presented tunable structure should be more applicable to space applications. In the case of space radiative cooling, the radiated power is a critical component in evaluating the performance of these designs. In high temperature situations where cooling is needed, the rectangular and cylindrical designs can radiate more power than the design without $Si_3N_4$ inclusions. In situations where cooling is not needed, the low cooling-power stretched states of all three designs reach similarly low levels. Thus, the inclusion of the $Si_3N_4$ structures improves the ability of these PDMS designs to respond to a dynamic temperature environment.

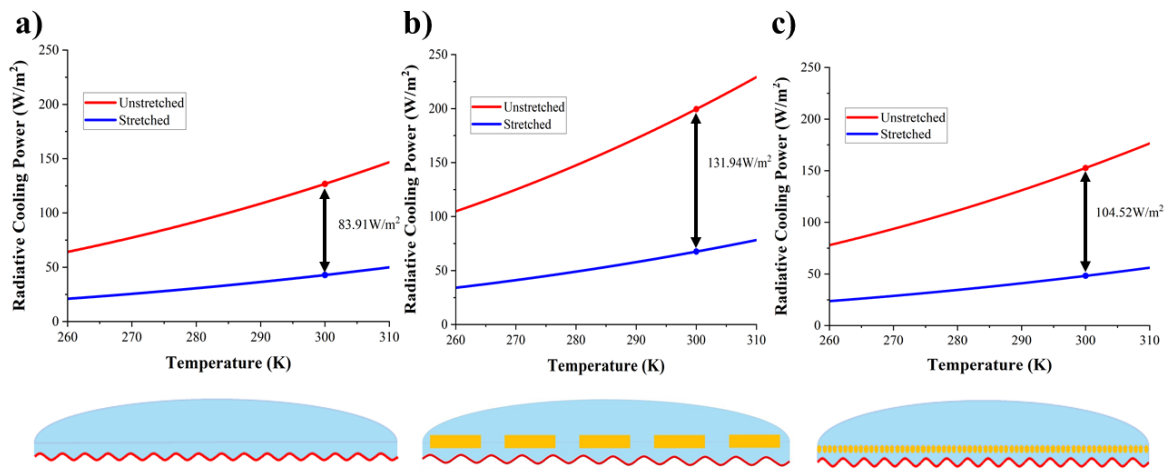

**Figure 7.** Radiated cooling power versus temperature for the PDMS over crumpled metal designs a) without $Si_3N_4$ inclusions, b) with rectangular $Si_3N_4$, and c) with cylindrical $Si_3N_4$.

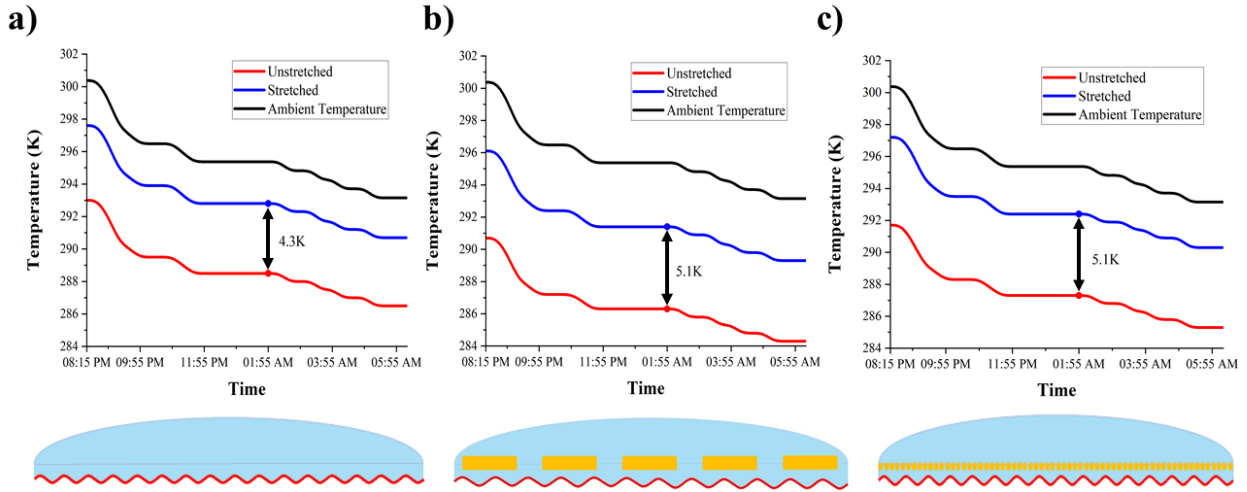

**Figure 8.** Computed hourly temperature plots or the PDMS over crumpled metal designs a) without $Si_3N_4$ inclusions, b) with rectangular $Si_3N_4$, and c) with cylindrical $Si_3N_4$ in a terrestrial nocturnal radiative cooling scenario.

## CONCLUSION

In conclusion, we have demonstrated high-performance mechanically tunable radiative cooling for extraterrestrial (satellites) and terrestrial (nocturnal) applications utilizing patterned dielectric ($Si_3N_4$) structures embedded within an elastic material (PDMS). The decrease in the thickness and increase in the periodicity due to the stretching serve as the main mechanisms that reduce the emissivity of the structures in the mid-IR leading to a tunable cooling power response. The addition of a crumpled metal substrate further boosts the tunability and cooling performance by increasing the reflectance of the substrate when stretched. Compared to previous stretchable radiative cooling structures, our designs are simpler to fabricate and provide higher radiated power combined with larger tunable response. In addition, the currently developed theoretical model can be applied to any tunable thermal emitter based on mechanically stretchable materials, since it can accurately calculate mechanical stretching and its effect in the electromagnetic and thermal response. Extending the capabilities of radiative cooling devices to provide dynamic tunability will substantially improve the practicability of this emerging renewable-energy technology to many useful and exciting extraterrestrial applications, such as geostatic satellites operating in the Earth's shadow, sun synchronous satellites, or terrestrial nocturnal applications for tunable radiative cooling in the night. Furthermore, the presented designs could feasibly be mounted on mechanical louvers [56] that could control and point the absorptive/emissive side of our design away from the sun, since they are either transparent or absorptive to the sun's radiation [47].

**Corresponding Author**

Christos Argyropoulos - Department of Electrical and Computer Engineering, University of Nebraska-Lincoln, Lincoln, Nebraska 68588, USA. Email: christos.argyropoulos@unl.edu

**Authors**

Andrew Butler - Department of Electrical and Computer Engineering, University of Nebraska-Lincoln, Lincoln, Nebraska 68588, USA.

## ACKNOWLEDGMENTS

This work was partially supported by the Office of Naval Research Young Investigator Program (ONR-YIP) (Grant No. N00014-19-1-2384), the National Science Foundation/EPSCoR RII Track-1: Emergent Quantum Materials and Technologies (EQUATE) (Grant No. OIA-2044049), and the NASA Nebraska Space Grant Fellowship.

# Supplementary Information

# Mechanically Tunable Radiative Cooling for Adaptive Thermal Control

Andrew Butler and Christos Argyropoulos[*]

Department of Electrical and Computer Engineering, University of Nebraska-Lincoln, Lincoln, Nebraska 68588, USA

*christos.argyropoulos@unl.edu

**Electromagnetic Simulation Details**

The finite element method (FEM) full-wave simulation software Comsol Multiphysics® was used to calculate the mid-IR emissivity of our various PDMS tunable structures. Port boundaries at the top and bottom edges of the geometry were used to generate incident transverse electric (TE) and tranverse magnetic (TM) polarized plane waves. The incident angle of the plane wave $\theta$ is defined with respect to the y-axis. The side boundaries are simulated by using periodic boundary conditions. The simulation parameters are summarized in Figure S1 using the rectangular $Si_3N_4$ structures as an example. The same setup was used for the simulations of the cylindrical $Si_3N_4$ designs. The structures were modelled in two dimensions since they are assumed to be very large in the z-direction. PDMS [1] and $Si_3N_4$ [2] were simulated using frequency dependent permittivity data taken from the literature.

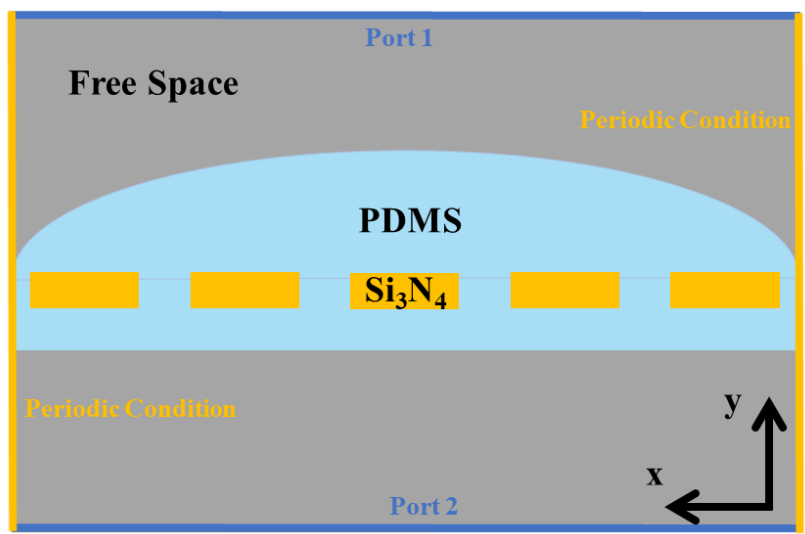

**Figure S1.** Schematic of the simulation domain.

The stretching of the structures was simulated by calculating the new dimensions of the structure when a 120% lateral strain is introduced in the x-direction. When stretched, dimensions in the x-direction are increased proportional to the strain. On the contrary, the dimensions in the y-direction are decreased proportional to the strain and the Poisson's Ratio of PDMS [3]. The exceptions to this effect are the dimensions of the $Si_3N_4$ structures which remain the same due to the high Young's Modulus of $Si_3N_4$ [4]. However, the spacing between the $Si_3N_4$ structures is increased with the strain. The dimension size in the x-direction after stretching can be calculated by using the formula:

$$x = x_0 + \epsilon * x_0, \quad (1)$$

where $x$ is the dimension in the x-direction such as the period of the PDMS or the period of the $Si_3N_4$ structures, $x_0$ is the size of the dimension before stretching, and $\varepsilon$ is the strain. The size of dimensions in the y-direction are similarly calculated by using the relation:

$$y = y_0 - \varepsilon * \upsilon * y_0, \quad (2)$$

where $y$ is the dimension in the y-direction such as the thickness of the flat PDMS or the height of the PDMS mound, $y_0$ is the size of the dimension before stretching, and $\upsilon$ is the Poisson's Ratio of PDMS. In our simulations, we use a maximum strain of 120% and a value of 0.48 for the Poisson's Ratio of PDMS [5,6].

S-parameter reflection and transmission coefficient calculations were performed to determine the mid-IR emissivity of our structures before and after stretching. The port boundaries were used to determine the total reflected and transmitted power flow through our structures. The power flow through each port was calculated as:

$$P = \int_C \vec{S} \cdot \vec{n} = \frac{1}{2} \int_C Re\{\vec{E} \times \vec{H}^*\} \cdot \vec{n}, \quad (3)$$

where $\vec{S}$ is the time averaged Poynting Vector, C is the curve of the port boundary, $\vec{n}$ is the normal vector, $\vec{E}$ is the electric field vector, and $\vec{H}$ is the magnetic field vector. The S-parameters were then computed by:

$$S_{11} = \sqrt{\frac{power\ reflected\ back\ to\ port\ 1}{power\ emitted\ from\ port\ 1}}, \quad (4)$$

$$S_{21} = \sqrt{\frac{power\ transmitted\ to\ port\ 2}{power\ emitted\ from\ port\ 1}}. \tag{5}$$

By using the Kirchhoff's law of thermal equilibrium, the emissivity $E$ is equal to the absorptance $A$ in the far field of a structure. The power that is not transmitted or reflected from the structure must be absorbed, therefore the absorptance is computed as: $A = 1 - |S_{11}|^2 - |S_{21}|^2$. Using this method, the full-wave simulations were performed twice for TM polarization, once for the stretched geometry and once for the unstretched geometry, to obtain the full spectral and angular emissivity of the presented structures. The same simulations were then repeated for TE polarization and the final emissivity used in the calculations of the cooling power was taken as the average of the two polarization results, since thermal emission is incoherent and contains both polarizations.

## Electric Field Distributions

Figures S2 and S3 show surface plots of the electric field enhancement distributions by using TM polarized incident radiation for the rectangular and cylindrical $Si_3N_4$ inclusions embedded in the PDMS designs, respectively. These plots help illustrate the physical mechanisms that enable the large differences in mid-IR emissivity between the stretched and unstretched states. At a wavelength of 9.2 μm, the emissivity is mainly caused by absorption in the PDMS layer. When stretched, the PDMS layer becomes thinner and absorbs less power. Similarly, at the wavelength of 10.6 μm, the emissivity of our structures is primarily caused by absorption in the $Si_3N_4$ structures. When stretched, the $Si_3N_4$ structures move farther apart, and more power is transmitted through the total system. This is most clearly illustrated in Figure S2.

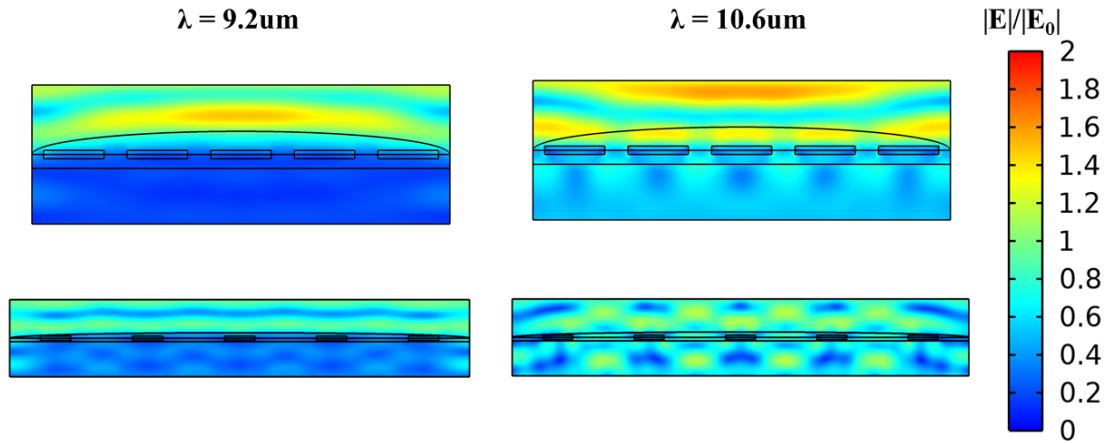

**Figure S2.** Surface plots of the electric field enhancement distributions for the rectangular $Si_3N_4$ inclusions in the PDMS structures at the peak absorption wavelengths of PDMS (9.2 μm) and $Si_3N_4$ (10.6 μm). The upper captions correspond to the unstretched structure while the lower captions to the stretched configuration.

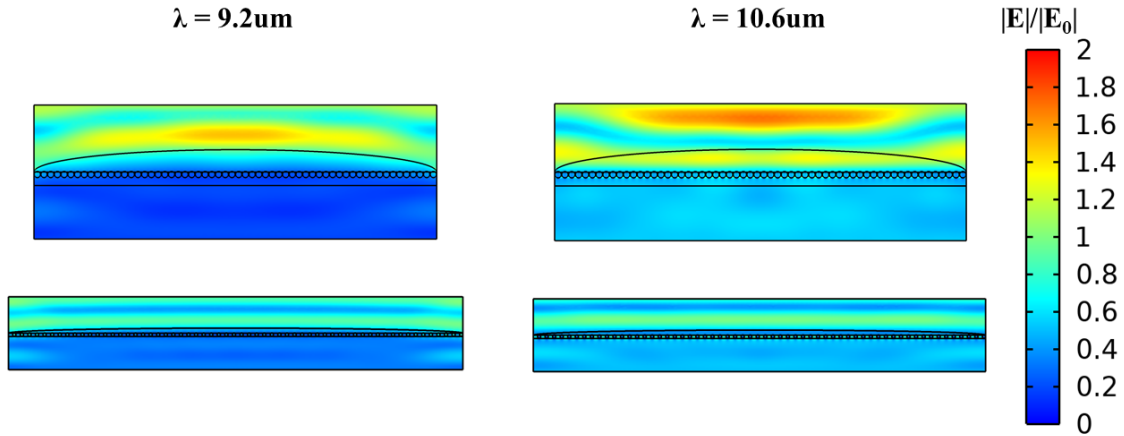

**Figure S3.** Surface plots of the electric field enhancement distributions for the cylindrical $Si_3N_4$ inclusions in the PDMS structures at the peak absorption wavelengths of PDMS (9.2 μm) and $Si_3N_4$ (10.6 μm). The upper captions correspond to the unstretched structure while the lower captions to the stretched configuration.

Figures S4 and S5 show the electric field enhancement distributions by using TM polarized incident radiation for the crumpled metal PDMS designs with rectangular and cylindrical $Si_3N_4$ inclusions, respectively. In these cases, the metallic layer prevents the transmission of light through the structure. Thus, the reduction in absorption/emissivity is due to an increase in reflection. The flattening of the metal as well as the increase in periodicity of the $Si_3N_4$ structures are the mechanisms by which the reflectance is increased. When the space between the $Si_3N_4$ structures is increased, an expanded portion of the metallic surface is exposed, leading to increased reflectance.

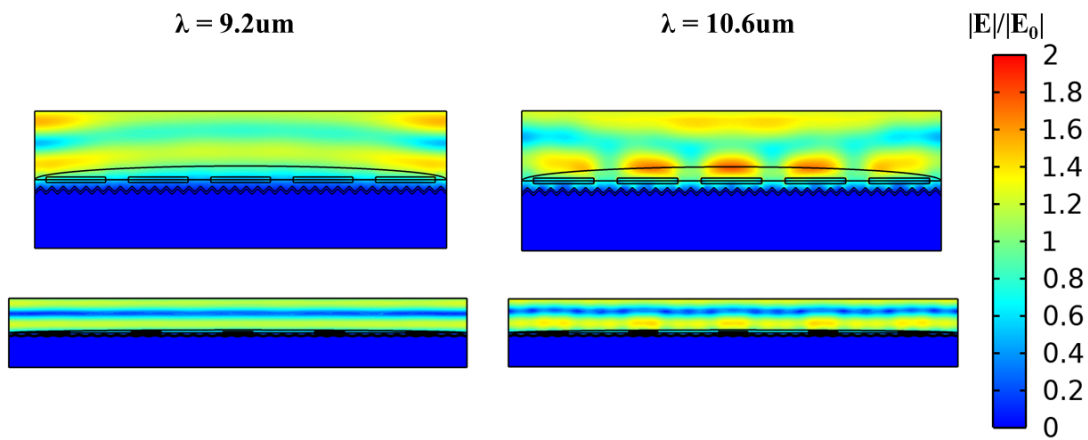

**Figure S4.** Surface plots of the electric field enhancement distributions for the rectangular $Si_3N_4$ inclusions within the crumpled metal PDMS structures at the peak absorption wavelengths of PDMS (9.2 μm) and $Si_3N_4$ (10.6 μm). The upper captions correspond to the unstretched structure while the lower captions to the stretched configuration.

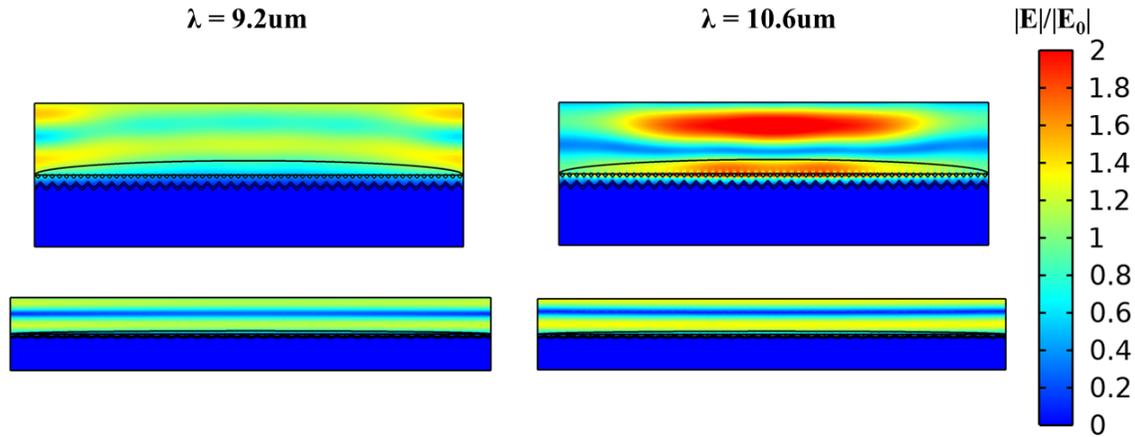

**Figure S5.** Surface plots of the electric field enhancement distributions for the cylindrical $Si_3N_4$ inclusions within the crumpled metal PDMS structures at the peak absorption wavelengths of PDMS (9.2 μm) and $Si_3N_4$ (10.6 μm). The upper captions correspond to the unstretched structure while the lower captions to the stretched configuration.

### $Si_3N_4$ Shape Effect versus Size Effect

The dimensions of the cylindrical $Si_3N_4$ structures are much smaller than the rectangular structures. The differences in the results of the rectangular structures versus the cylindrical structures in Fig. 6 in the main paper can either be due to the shape of the structures or the large difference in size. To determine the main cause of the difference in the results, simulations were carried out for the design with the metal substrate, replacing the cylindrical structures with square shapes of the same size such that the width and thickness of the squares were equal to the diameter of the cylindrical structures used in the main paper. Figure S6a shows the emissivity spectra for these square shapes (highlighted as the blue area in the inset schematic). The emissivity near λ = 11μm is slightly higher for this case than for the cylindrical case (Figure S6b). While the dimensions of the square shaped structures are the same as the cylindrical structures, the area of the square shapes is larger. Figure S6c shows the emissivity spectra of the design with square shaped structures that are slightly smaller than the cylindrical structures (sized such that the squares are inscribed on the circles and highlighted as the blue area in the inset schematic). In this case, the emissivity near λ = 11μm is smaller than both other cases. This indicates that the emissivity due to the $Si_3N_4$ structures is mainly dependent on the size (area) of the structures and not their shapes.

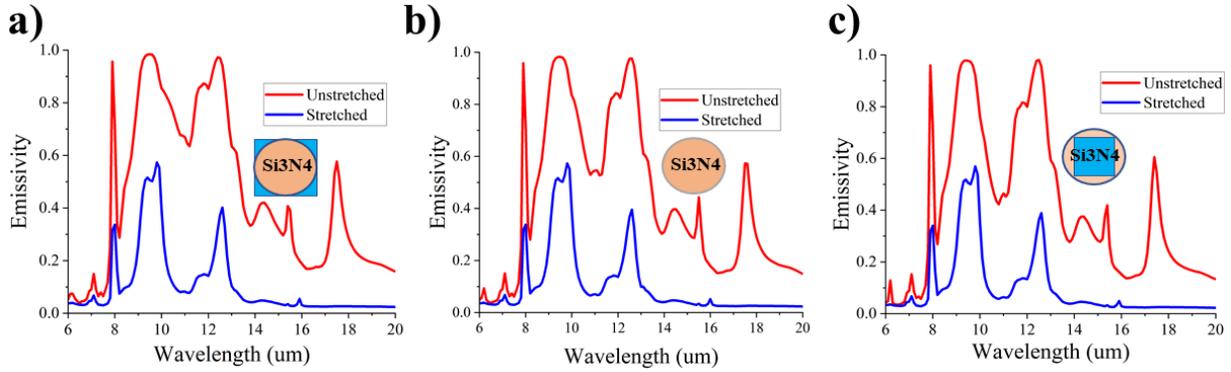

**Figure S6.** Emissivity spectra (TM polarized) of the PDMS device with a metallic substrate and (a) square $Si_3N_4$ structures of the same size as the original cylindrical design, (b) the cylindrical structures from the main paper, and (c) square shaped structures that are slightly smaller than the original cylindrical structures.

**Effect of $Si_3N_4$ Dimensions**

The dimensions of the $Si_3N_4$ structures have a minor effect on the performance of the designs. In this section we show the dependence of the emissivity on the geometric parameters of the $Si_3N_4$ structures. We begin first with the thickness of the rectangular structures and the radius of the cylindrical structures. Figure S7a shows the average emissivity over the 9-13μm wavelength range versus the thickness of the rectangular structures while Figure S7b shows average emissivity versus the radius of the cylindrical structures. The metallic substrate was included in the simulation and all other dimensions including the periodicities, rectangular width, and PDMS thicknesses were kept the same as the designs presented in Figure 2 of the main paper.

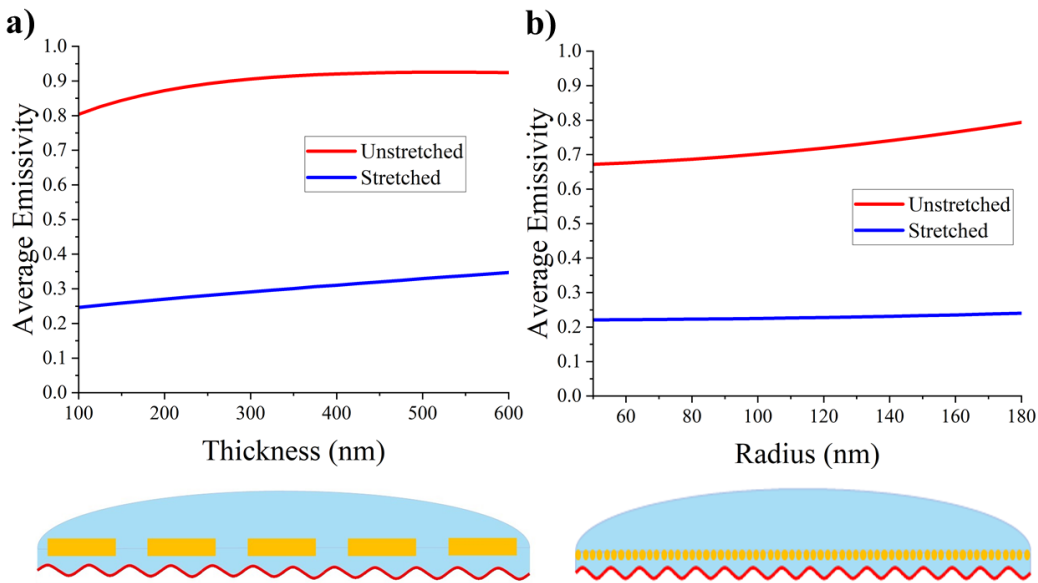

**Figure S7.** Average emissivity (TM polarized) between 9μm and 13μm versus (a) the thickness of the rectangular structures and (b) the radius of the cylindrical $Si_3N_4$.

In general, the average emissivity slightly increases with the size of the Si$_3$N$_4$ structures. The flattening of the curve for the rectangular case indicates that the results are not sensitive to the thickness. Hence, fabrication imperfections will have little effect on the results. The effect of the periodicity and width of the rectangular structures is shown in Figure S8. Here, the average emissivity is plotted for two different periodicities as a function of the fill fraction of the structures. The fill fraction is defined as the width of the structures divided by the total periodicity. For these simulations, the metal substrate was included and a value of 600nm was used for the rectangular thickness. The possible values of the periodicity are limited by the dimensions of the PDMS. The possible periodicities are first limited by the periodicity of the PDMS and should be constrained to values that divide the PDMS periodicity evenly to avoid having fractions of a unit cell embedded in the PDMS. Second, since the size and thickness of the Si$_3$N$_4$ structures remain unchanged during stretching, the periodicity, width, and thickness cannot be greater than the thickness of the PDMS when stretched, otherwise the PDMS would rupture and crack and the Si$_3$N$_4$ structures would break through the PDMS when stretched. At small periodicities, more rectangular structures are located at the edge of the device where the curved PDMS gets thinner, and it is more likely that the Si$_3$N$_4$ structures will break through the PDMS. With these restrictions we limit our investigation to two larger periodicities of 9μm and 15μm.

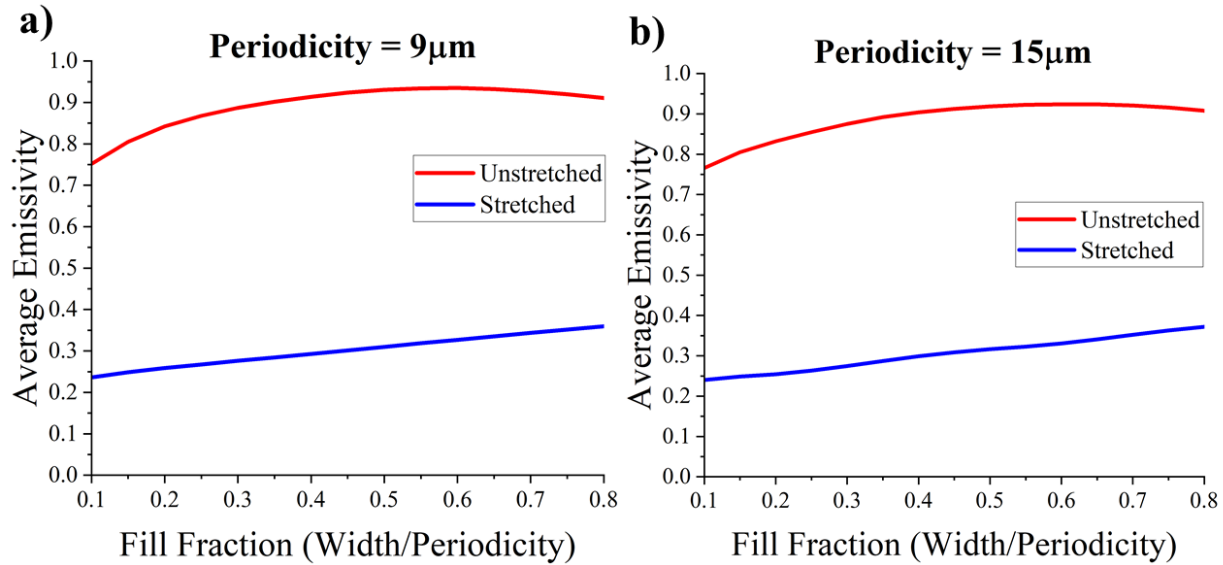

**Figure S8.** Average emissivity (TM polarized) between 9μm and 15μm versus the rectangular Si$_3$N$_4$ fill fraction for periodicities of (a) 9μm and (b) 15μm.

Once again, the general trend shows that larger dimensions of Si$_3$N$_4$ lead to slightly higher emissivity. The difference between the two periodicities is small and once again the curve flattens at higher fill fractions. Thus, variations in the width of the rectangular structures due to fabrication errors will have little effect on the overall performance of the device. Similar simulations were carried out for the cylindrical design to show its dependence on the periodicity. The same constraints apply for the cylindrical design and limit the possible periodicities. Since the effects of the radius have already been demonstrated before in Figure S7b, Figure S9 shows the emissivity spectra for three different periodicities of the cylindrical design. For these simulations, a radius of 180nm was used and the metal substrate was included.

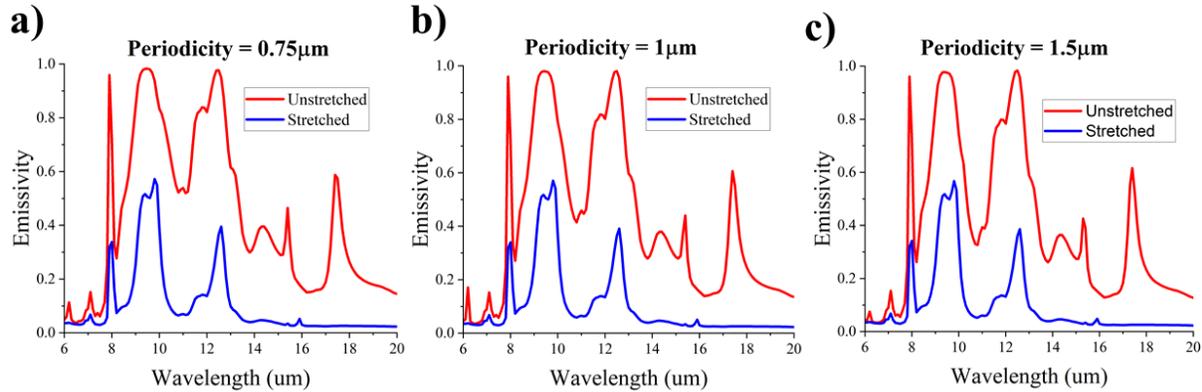

**Figure S9.** Emissivity spectra (TM polarized) of the cylindrical design for periodicities of (a) 0.75μm, (b) 1μm, and (c) 1.5μm.

The only notable effect of the cylindrical periodicity occurs near $\lambda = 11$μm where the emissivity slightly increases as the periodicity decreases. For the cylindrical case, a smaller periodicity means that the gap between the cylinders is smaller and thus the fill fraction is larger. Hence, this behavior is similar to that of the rectangular design presented before in Figure S8. Overall, the general trend shows that the emissivity slightly increases with the size of the $Si_3N_4$ structures, though at a certain point, the size effect diminishes. Since the $Si_3N_4$ structures are primarily responsible for the induced emissivity near $\lambda = 11$μm, small variations in dimensions that could occur during fabrication would only result in a miniscule effect on the overall performance of the device.

## Device Behavior in the Solar Spectrum

For many radiative cooling applications, sunlight plays a major factor in the device performance as a large detrimental heating source. Here we demonstrate the absorption spectra of our designs in the solar spectrum range. Figure S10 shows the solar absorption spectra of the rectangular designs with and without the metal substrate while Figure S11 shows the solar absorption spectra for the cylindrical designs.

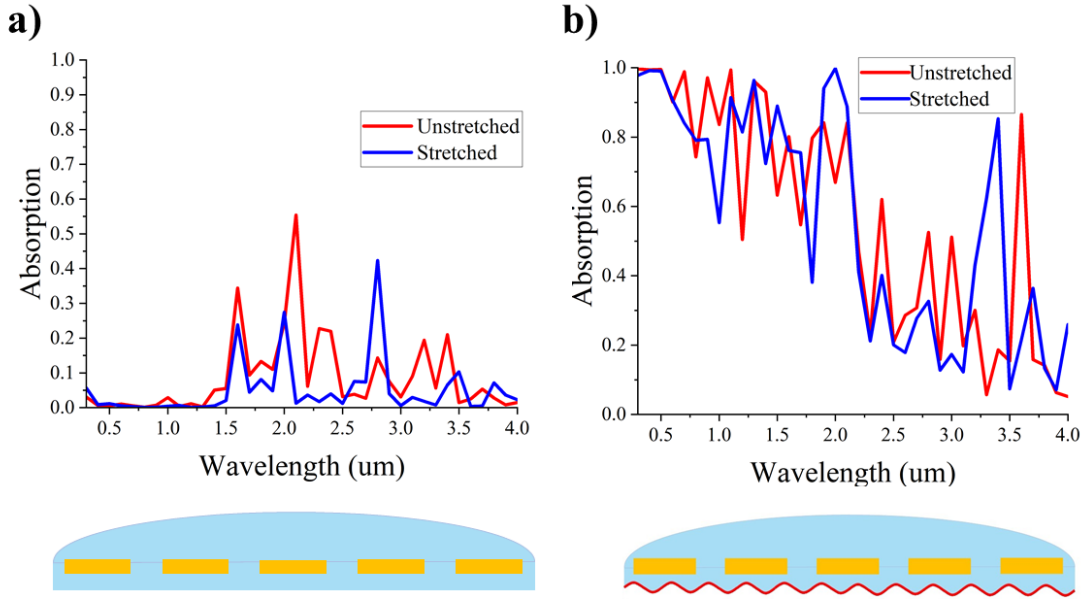

**Figure S10.** Solar absorption spectra (TM polarized) of the rectangular designs (a) without and (b) with metal substrate.

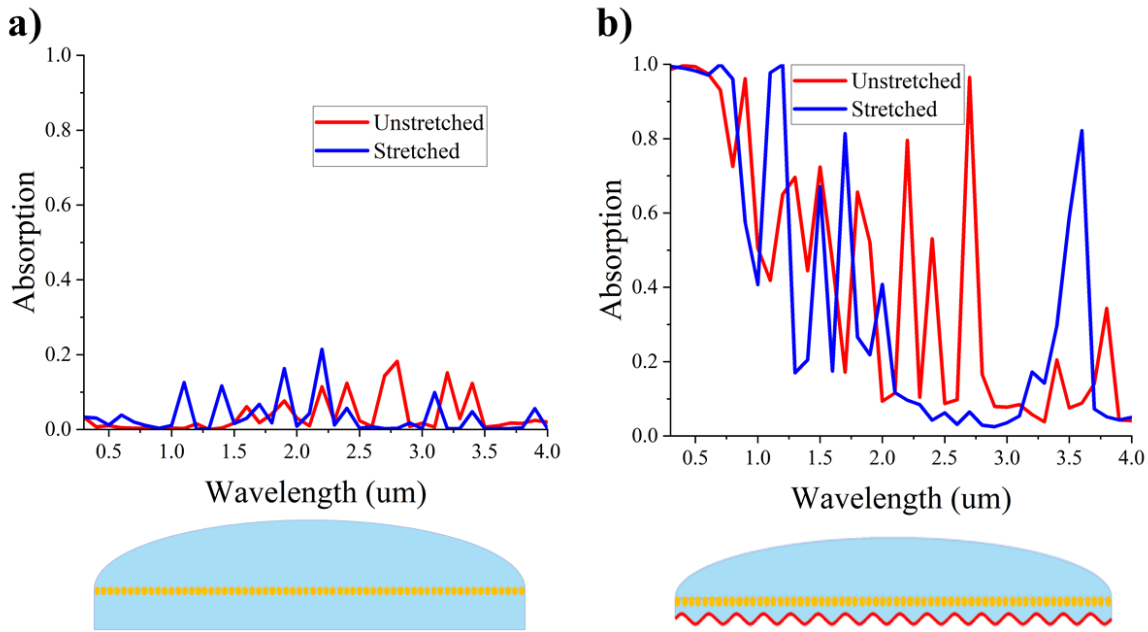

**Figure S11.** Solar absorption spectra (TM polarized) of the cylindrical designs (a) without and (b) with metal substrate.

It is clear that the inclusion of the crumpled metal substrate causes high absorption of solar light. Whereas for the designs without metal, the absorption is much lower due to being mostly transparent to solar light. While this behavior does somewhat limit the space applications of the presented designs, there are still useful applications for these devices on geostatic satellites operating in the Earth's shadow or sun synchronous satellites. In the Earth's shadow, no sunlight reaches the satellite and so conserving energy becomes an important concern since the solar panels are no longer generating electrical power. The ability to reduce the amount of power radiating from the satellite becomes an important benefit of our designs. It is also feasible that the radiative cooling devices could be mounted and controlled by using mechanical louvers [7], such that the absorptive/emissive side faces away from the sun while backed with a reflective metal surface.